%% file: Main.tex
\documentclass{article}
\usepackage{spconf,amsmath,graphicx}

\title{Robust Autoencoders for Collective Corruption Removal}
\name{Taihui Li $^{1}$ \quad Hengkang Wang $^{1}$ \quad Peng Le $^{1}$\quad Xian'e Tang $^{2}$\quad Ju Sun $^{1}$ \thanks{HW, PL, and JS are partially supported by NSF CMMI 2038403.}}

\address{${^1}$ Computer Science and Engineering, University of Minnesota, Minneapolis, USA \\
${^2}$ Computer Science, University of Minnesota Duluth, Duluth, USA}

\input{ju_math}
\usepackage{booktabs}
\usepackage{multirow}
\usepackage{makecell}

\begin{document}
%
\maketitle
\begin{abstract}
\input{sections/Sec0_Abstract}
\end{abstract}
%

\input{sections/Sec1_Introduction}
\input{sections/Sec2_Preliminaries}
\input{sections/Sec3_Method}

\input{sections/Sec4_Exp}
\input{sections/Sec6_Dis}
\bibliographystyle{IEEEbib}
\bibliography{Main}

\end{document}

%% file: ju_math.tex
\usepackage{graphicx,amsmath,amsfonts,amsthm,amscd,amssymb,bm,url,color,latexsym}
\usepackage{physics}
\allowdisplaybreaks
\usepackage[capitalize,nameinlink]{cleveref}

\renewcommand{\mathbf}{\boldsymbol}

\newcommand{\mb}{\mathbf}

\newcommand{\bb}{\mathbb}

\newcommand{\R}{\bb R}

\newcommand{\paren}{\pqty}

\newcommand{\wh}{\widehat}



%% file: sections/Sec0_Abstract.tex
Robust PCA is a standard tool for learning a linear subspace in the presence of sparse corruption or rare outliers. What about robustly learning manifolds that are more realistic models for natural data, such as images? There have been several recent attempts to generalize robust PCA to manifold settings. In this paper, we propose $\ell_1$- and \textbf{scaling-invariant} $\ell_1/\ell_2$-robust autoencoders based on a surprisingly compact formulation built on the intuition that deep autoencoders perform manifold learning. We demonstrate on several standard image datasets that the proposed formulation significantly outperforms all previous methods in collectively removing sparse corruption, without clean images for training. Moreover, we also show that the learned manifold structures can be generalized to unseen data samples effectively. 

%% file: sections/Sec1_Introduction.tex
\section{Introduction}\label{sec:introduction}
Principal component analysis (PCA) is a widely used statistical method for dimensionality reduction. However, it is known to be sensitive to data corruption and outliers~\cite{HuberRonchetti2009Robust}. Over the last decade, several robust formulations of PCA have been proposed, and they are guaranteed to be robust against sparse additive corruption or rare
outliers (e.g., \cite{candes2011robust,chandrasekaran2011rank}). These methods take a common geometric view of the problem: robust PCA (RPCA) tries to find a best-fit subspace for a point cloud contaminated by gross errors. They differ by how to encode the prior knowledge that the desired subspace is low-dimensional and that the gross errors are sparse. 

Consider the sparse-additive-corruption model: let $\mb x_1,\dots, \allowbreak \mb x_m \allowbreak \in \allowbreak\R^n$ be a set of data points lying in a low-dimensional subspace and write $\mb X = [\mb x_1, \dots, \mb x_m] \in \R^{n \times m}$. We hope to recover $\mb X$ from the sparsely corrupted version 
\begin{align}  \label{eq:data_model}
\wh{\mb X} = \mb X + \mb C,\quad \text{(data model)}
\end{align}
where $\mb C$ is a sparse matrix containing very few nonzero entries with potential large magnitude. To solve this problem, a seminal convex formulation is~\cite{candes2011robust}:
\begin{equation}
\label{eq:rpca}
\begin{aligned}
	\min_{\mb X, \mb C} & \quad \norm{\mb X}_{\ast}  + \lambda \norm{\mb C}_{1} \\
\mathrm{s.t.}   & \quad \wh{\mb X} = \mb X + \mb C 
\end{aligned}. 
\quad \text{(convex RPCA)}
\end{equation}
Here $\norm{\cdot}_{\ast}$ is the nuclear norm, i.e., the sum of singular values, serving as a surrogate for matrix rank, and $\norm{\cdot}_1$ is a surrogate for sparsity. So intuitively, if $\lambda$ is chosen appropriately, solving \cref{eq:rpca} can recover $\mb X$ and $\mb C$ simultaneously. 

\emph{What about when the data $\mb X$ lie on low-dimensional manifolds, natural generalization of subspaces?} For example, in computer vision, a collection of images of the same object/scene under varying illumination, deformation, and viewpoint is typically assumed to lie on a low-dimensional manifold~\cite{lu1998image,belhumeur1998set}. However, shadows, occlusions, and sensory defects that are frequently present in practical vision datasets often do not correspond to any explanatory factors of interest and so can be treated as corruption, gross errors, and outliers to the data manifolds. Therefore, representation learning for vision data, which is essential for virtually all high-level vision tasks, can be regarded as robust learning of manifold structures in the presence of corruption, errors, and outliers. 

We focus on robust manifold learning against \emph{sparse corruption} as a first step. Sparse corruption is common in applications and can model useful signals (e.g., moving objects in video surveillance~\cite{candes2011robust}) or simply noise (e.g., salt-and-pepper noise \cite{wang_early_2021,li_self-validation:_2021} due to sensor failure in image acquisition or errors during transmission~ \cite{li_impulse_2021,yuan_ell_2019,adam_combined_2021}). Our main idea is to find a nonlinear mapping (\emph{forward mapping}) that maps the high-dimensional noisy data $\wh{\mb X}$ near an unknown manifold to low-dimensional representations, and an ``inverse" mapping (\emph{backward mapping}) that maps the low-dimensional representations back to the origin space which hopefully recovers the clean data $\mb X$. Through this 
\emph{forward-backward} mapping process, we hope to mitigate the sparse corruption $\mb C$ and recover the clean data $\mb X$.

%% file: sections/Sec2_Preliminaries.tex
\section{Prior Work}\label{subsec:generalization_others}


In this section, we review three recent methods on robust manifold learning and highlight their limitations.

\emph{\textbf{Nonlinear robust PCA (NRPCA)}} 
\cite{lyu2019manifold} performs localized RPCA for sparsely corrupted data points from a manifold. It relies on a basic geometric fact: for any point on a manifold, it neighboring set of the manifold can be well approximated by a low-dimensional subspace (i.e., the tangent space in differential geometry). So each noisy data point and its neighboring data points can be thought of sparsely corrupted data points from a low-dimensional subspace, i.e., the RPCA problem. \cite{lyu2019manifold} performs repeated RPCA around all the data points based on a modified form of \cref{eq:rpca}. The main limitation of this method is that it does not perform manifold learning, but only restricts itself to corruption removal. 



 \emph{\textbf{Robust autoencoder (RAE-CHA)}} 
 \cite{chalapathy2017robust} draws inspiration from both shallow autoencoders and a nonconvex formulation of RPCA to arrive at 
\begin{equation}\label{eq:another_rda}
\begin{aligned}
\min_{\mb A, \mb B, \mb C} \;  \norm{\wh{\mb X} - \mb B\sigma(\mb A (\wh{\mb X} )) - \mb C}_{F}^2 +  \\
 \quad \quad \lambda_1 \paren{\norm{\mb A}_F^2 + \norm{\mb B}_F^2}
+ \lambda_2 \norm{\mb C}_1.
\end{aligned}
\; (\text{RAE-CHA})
\end{equation}
Here, $\mb B\sigma(\mb A (\cdot))$ is a nonlinear autoencoder with a one-layer encoder and a one-layer decoder, and $\mb C$ accounts for the sparse corruption. The regularization term $\norm{\mb A}_F^2 + \norm{\mb B}_F^2$ encourages low-complexity $\mb A$ and $\mb B$, and  has a direct  connection to the nuclear norm\footnote{The variational identity holds: for any matrix $\mb M$, $\norm{\mb M}_\ast \allowbreak = \allowbreak \min_{\mb U, \mb V}\; \allowbreak \frac{1}{2} \allowbreak(\norm{\mb U}_F^2 \allowbreak + \allowbreak \norm{\mb V}_F^2)$~\cite{srebro2004maximum}. This form of regularization has been popular in nonconvex formulations of RPCA, see, e.g.,~\cite{li2019non}. }. A major problem is that the shallow autoencoder $\mb B\sigma(\mb A (\cdot))$ cannot capture high-complexity structures in practical data, and it is not clear how to generalize the regularization term $\norm{\mb A}_F^2 + \norm{\mb B}_F^2$ when replacing it with a deep autoencoder.

\emph{\textbf{Robust deep autoencoder (RDA)}} Zhou and Paffenroth~\cite{zhou2017anomaly} takes advantage of deep autoencoders for manifold learning and proposes a formulation for robust manifold learning:
\begin{equation}\label{eq:kdd_rda}
\begin{aligned}
	\min_{\mb X, \mb C, \mb W, \mb V} &\; \norm{\mb X - g_{\mb V} \circ f_{\mb W}  \paren{\mb X}}_{F}  + \lambda \norm{\mb C}_{1} \; \\&s.t.\;\quad \wh{\mb X} = \mb X + \mb C
\end{aligned}
(\text{RDA}).
\end{equation}
Here, $\mb X$ and $\mb C$ account for the clean manifold data and the sparse corruption, respectively. $g_{\mb V} \circ f_{\mb W}\paren{\cdot}$ is a deep autoencoder. They propose a variant of the alternating-direction method of multipliers (ADMM)~\cite{boyd2011distributed} to solve the constrained problem. But in order to deal with unseen data, one needs to freeze $\mb V$ and $\mb W$ and solve for $\mb X$ and $\mb C$ based on \cref{eq:kdd_rda}, which entails nontrivial numerical optimization.

%% file: sections/Sec3_Method.tex
\section{Our Formulation}\label{sec:method}

\subsection{Autoencoders for Manifold Learning}  \label{subsec:pca}

For a data matrix $\mb X \in \R^{n \times m}$, where \emph{each column} is a data point, linear autoencoder learns linear mappings $\mb A \in \R^{p \times n}$ and $\mb B \in \R^{n \times p}$ so that
\begin{align} \label{eq:linear_ae}
\min_{\mb A, \mb B}\;  \norm{\mb X - \mb B \mb A \mb X}_{F}^2.  \quad (\text{linear autoencoder})
\end{align}
Typically $p < n$, and so $\mb A$ maps each data point in $\mb X$ to a low-dimensional representation (or ``code") and $\mb B$ maps the representation back to the original space. It is well understood that under mild conditions on $\mb X$, solving the formulation in \cref{eq:linear_ae} finds a $\mb B$ spanning the $p$-dimensional subspace that best approximates the column space of $\mb X$~\cite{baldi1989neural}. In other words, an optimal $\mb B$ spans the $p$-dimensional principal subspace of $\mb X$. In this sense, linear autoencoder effectively performs PCA or subspace learning, which was recognized in the early days since its invention~\cite{baldi1989neural}.

The geometric view of linear autoencoder has inspired people to generalize it for learning nonlinear (manifold) structures in practical data. The idea is simple: generalizing the linear mappings $\mb A$ and $\mb B$ into nonlinear mappings, which are often represented by deep neural networks (DNNs):
\begin{align}
\label{eq:deep_ae}
 \min_{\mb W, \mb V} \; \norm{\mb X - g_{\mb V} \circ f_{\mb W}  \paren{\mb X}}_{F}^2 \quad (\text{deep autoencoder}).
\end{align}
Here $f_{\mb W}$ generalizes $\mb A$ and $g_{\mb V}$ generalizes $\mb B$.  $f_{\mb W}$ and $g_{\mb V}$ are DNNs parameterized by $\mb W$ and $\mb V$, respectively.



\begin{table*}[!htpb]
\caption{The collective corruption removal performance of Baseline,
RDA~\cite{zhou2017anomaly}, NRPCA~\cite{lyu2019manifold}, RAE-CHA~\cite{chalapathy2017robust}, $\ell_1$-RAE (ours), $\ell_1/\ell_2$-RAE (ours). The best PSNRs are colored as \textcolor{red}{red}; the best SSIMs are colored as \textcolor{blue}{blue}.}
\vspace{-3mm}
\label{tab:given}
\begin{center}
\setlength{\tabcolsep}{0.9mm}{
\begin{tabular}{c c c c c c c c| c c c c c c}
&
& 
\multicolumn{6}{c}{\scriptsize{PSNR $\uparrow$}} &
\multicolumn{6}{c}{\scriptsize{SSIM $\uparrow$}}
\\
\hline

\hline
\scriptsize{Dataset}

& \scriptsize{Corr.}

& \scriptsize{Baseline}
& \scriptsize{RDA}
& \scriptsize{NRPCA}
& \scriptsize{RAE-CHA}
& \scriptsize{$\ell_1$-RAE}
& \scriptsize{$\ell_1/\ell_2$-RAE}

& \scriptsize{Baseline}
& \scriptsize{RDA}
& \scriptsize{NRPCA}
& \scriptsize{RAE-CHA}
& \scriptsize{$\ell_1$-RAE}
& \scriptsize{$\ell_1/\ell_2$-RAE}
\\
\hline

\hline
\multirow{4}{*}{\rotatebox{90}{\scriptsize{MNIST}}}

&\scriptsize{10\%}

&\scriptsize{13.23}
&\scriptsize{18.55}
&\scriptsize{17.42}
&\scriptsize{21.57}
&\scriptsize{24.38}
&\scriptsize{\textcolor{red}{24.46}}

&\scriptsize{0.60}
&\scriptsize{0.76}
&\scriptsize{0.61}
&\scriptsize{0.89}
&\scriptsize{\textcolor{blue}{0.96}}
&\scriptsize{\textcolor{blue}{0.96}}
\\
\cline{2-14}

&\scriptsize{20\%}

&\scriptsize{10.19}
&\scriptsize{15.85}
&\scriptsize{15.77}
&\scriptsize{19.58}
&\scriptsize{23.04}
&\scriptsize{\textcolor{red}{23.18}}

&\scriptsize{0.46}
&\scriptsize{0.57}
&\scriptsize{0.60}
&\scriptsize{0.80}
&\scriptsize{0.94}
&\scriptsize{\textcolor{blue}{0.95}}
\\
\cline{2-14}

&\scriptsize{30\%}
&\scriptsize{8.42}
&\scriptsize{12.78}
&\scriptsize{13.85}
&\scriptsize{17.09}
&\scriptsize{\textcolor{red}{20.57}}
&\scriptsize{20.40}

&\scriptsize{0.37}
&\scriptsize{0.41}
&\scriptsize{0.53}
&\scriptsize{0.64}
&\scriptsize{\textcolor{blue}{0.91}}
&\scriptsize{\textcolor{blue}{0.91}}
\\
\cline{2-14}

&\scriptsize{40\%}
&\scriptsize{7.16}
&\scriptsize{10.86}
&\scriptsize{11.93}
&\scriptsize{14.12}
&\scriptsize{\textcolor{red}{16.47}}
&\scriptsize{15.25}

&\scriptsize{0.30}
&\scriptsize{0.32}
&\scriptsize{0.44}
&\scriptsize{0.49}
&\scriptsize{\textcolor{blue}{0.80}}
&\scriptsize{0.75}
\\
\hline

\hline

\multirow{4}{*}{\rotatebox{90}{\scriptsize{CIFAR10}}}

&\scriptsize{10\%}
&\scriptsize{15.06}
&\scriptsize{22.41}
&\scriptsize{25.38}
&\scriptsize{21.67}
&\scriptsize{36.36}
&\scriptsize{\textcolor{red}{38.20}}

&\scriptsize{0.47}
&\scriptsize{0.74}
&\scriptsize{0.87}
&\scriptsize{0.75}
&\scriptsize{0.98}
&\scriptsize{\textcolor{blue}{0.99}}
\\
\cline{2-14}

&\scriptsize{20\%}
&\scriptsize{12.05}
&\scriptsize{19.47}
&\scriptsize{23.24}
&\scriptsize{20.86}
&\scriptsize{31.23}
&\scriptsize{\textcolor{red}{36.33}}

&\scriptsize{0.31}
&\scriptsize{0.62}
&\scriptsize{0.79}
&\scriptsize{0.71}
&\scriptsize{0.96}
&\scriptsize{\textcolor{blue}{0.98}}
\\
\cline{2-14}

&\scriptsize{30\%}
&\scriptsize{10.28}
&\scriptsize{16.72}
&\scriptsize{22.71}
&\scriptsize{20.34}
&\scriptsize{27.38}
&\scriptsize{\textcolor{red}{31.88}}

&\scriptsize{0.22}
&\scriptsize{0.49}
&\scriptsize{0.77}
&\scriptsize{0.69}
&\scriptsize{0.90}
&\scriptsize{\textcolor{blue}{0.96}}
\\
\cline{2-14}

&\scriptsize{40\%}
&\scriptsize{9.03}
&\scriptsize{14.66}
&\scriptsize{19.25}
&\scriptsize{19.97}
&\scriptsize{22.74}
&\scriptsize{\textcolor{red}{25.46}}

&\scriptsize{0.16}
&\scriptsize{0.37}
&\scriptsize{0.63}
&\scriptsize{0.68}
&\scriptsize{0.77}
&\scriptsize{\textcolor{blue}{0.88}}
\\
\hline

\hline

\multirow{4}{*}{\rotatebox{90}{\scriptsize{CIFAR100}}}

&\scriptsize{10\%}
&\scriptsize{14.96}
&\scriptsize{22.45}
&\scriptsize{24.94}
&\scriptsize{21.83}
&\scriptsize{34.76}
&\scriptsize{\textcolor{red}{36.38}}

&\scriptsize{0.46}
&\scriptsize{0.73}
&\scriptsize{0.86}
&\scriptsize{0.75}
&\scriptsize{0.98}
&\scriptsize{\textcolor{blue}{0.99}}
\\
\cline{2-14}

&\scriptsize{20\%}
&\scriptsize{11.94}
&\scriptsize{19.71}
&\scriptsize{22.91}
&\scriptsize{20.81}
&\scriptsize{30.83}
&\scriptsize{\textcolor{red}{35.89}}

&\scriptsize{0.29}
&\scriptsize{0.62}
&\scriptsize{0.77}
&\scriptsize{0.71}
&\scriptsize{0.96}
&\scriptsize{\textcolor{blue}{0.99}}
\\
\cline{2-14}

&\scriptsize{30\%}
&\scriptsize{10.18}
&\scriptsize{17.28}
&\scriptsize{22.65}
&\scriptsize{20.17}
&\scriptsize{26.93}
&\scriptsize{\textcolor{red}{31.26}}

&\scriptsize{0.21}
&\scriptsize{0.50}
&\scriptsize{0.75}
&\scriptsize{0.69}
&\scriptsize{0.92}
&\scriptsize{\textcolor{blue}{0.97}}
\\
\cline{2-14}

&\scriptsize{40\%}
&\scriptsize{8.93}
&\scriptsize{15.23}
&\scriptsize{19.01}
&\scriptsize{19.53}
&\scriptsize{22.93}
&\scriptsize{\textcolor{red}{25.42}}

&\scriptsize{0.15}
&\scriptsize{0.39}
&\scriptsize{0.59}
&\scriptsize{0.64}
&\scriptsize{0.82}
&\scriptsize{\textcolor{blue}{0.90}}
\\
\hline

\hline
\end{tabular}
}
\end{center}
\end{table*}

\subsection{Robust Manifold Learning: the Natural Way}
Deep autoencoders perform manifold learning in a good sense, as discussed above. But how to achieve robustness against sparse corruption? A natural way is again to generalize formulations for the linear case, i.e., RPCA, to the manifold setting. If we were to generalize \cref{eq:rpca}, the only change we need to make is to encode the information that the clean data lies on manifolds, instead of being low-rank. So the nuclear norm term $\norm{\mb X}_\ast$ can be replaced with an autoencoder-like objective, e.g.,
\begin{align}
\norm{\mb X - g_{\mb V} \circ f_{\mb W}\paren{\mb X}}_F,
\end{align}
which leads exactly to the RDA formulation in \cref{eq:kdd_rda}. But as we alluded to above, the ADMM-based numerical method proposed in \cite{zhou2017anomaly} seems suboptimal, its generalization to unseen data entails nontrivial computation, and its empirical performance is off the simpler method that we propose below.

\subsection{Robust Manifold Learning: Our Simple Way}  \label{sec:rml_simple}
Our formulation is stunningly simple:
\begin{align}\label{eq:l1_ae}
 \min_{\mb W, \mb V} \; \norm{\wh{\mb X} - g_{\mb V} \circ f_{\mb W}  (\wh{\mb X})}_{1}. \quad (\text{$\ell_1$-RAE})
\end{align}
Here we hope that $ g_{\mb V} \circ f_{\mb W}  (\wh{\mb X})$ recovers the clean data $\mb X$, hence $\wh{\mb X} - g_{\mb V} \circ f_{\mb W}  (\wh{\mb X})$ is the sparse corruption regularized by the sparse regularizer $\norm{\cdot}_1$.

One might think that this is simply a matter of simply changing the loss in the deep autoencoder objective from $\norm{\cdot}_F^2$ to $\norm{\cdot}_1$. But that this formulation can work is not intuitive at all. We can think of a corresponding formulation for the linear case, i.e., RPCA:
\begin{align} \label{eq:rpca_ncvx}
 \min_{\mb B \in \R^{n \times p}, \mb A \in \R^{p \times n}} \; \norm{\wh{\mb X} - \mb B \mb A  \wh{\mb X}}_{1}.
\end{align}
Assume $\mb X$ is exactly rank-$p$. Now if $\mb B \mb A \wh{\mb X}$ were to recover the clean data $\mb X$, $\mb B$ must be column full-rank and $\mb A$ must be row full-rank. Let $\mathrm{row}\paren{\cdot}$ denote the row space.  Then, $\mathrm{row}\paren{\mb X} = \mathrm{row}(\mb B \mb A \wh{\mb X}) = \mathrm{row}(\mb A \wh{\mb X}) \subset \mathrm{row}(\wh{\mb X})$. Since typically sparse corruption can substantially alter the row space, it is very unlikely that $\mathrm{row}\paren{\mb X} \subset \mathrm{row}(\wh{\mb X})$. Therefore, the formulation in \cref{eq:rpca_ncvx} in general cannot solve the RPCA problem. Here, the possibility that $\ell_1$-RAE might work for nonlinear data relies on the nonlinear mappings $f_{\mb W}$ and $g_{\mb V}$---these defy the linear algebraic argument we made for the linear case.



Furthermore, to promote sparsity, $\ell_1$ norm is a standard choice, but not necessarily the best in practice. An obvious drawback of the $\ell_1$ norm is that it is \textbf{scale-sensitive}, i.e., scaling up or down the input matrix will correspondingly scale up or down the output value---this is not a desired property when approximating the $\ell_0$ norm, which is scale-invariant. To achieve invariance, one can use ratios of norms to remove the effect of scaling. For example, $\ell_1/\ell_2$ has been a popular soft measure of sparsity (vs. the hard measure $\ell_0$ which directly counts the number of nonzeros)~\cite{HoyerNon}. Its advantage over $\ell_1$ in terms of sparse recovery in the compressive sensing setting and image deblurring has recently been established~\cite{rahimi2019scale,zhuang_blind_2022}. So we also propose an $\ell_1/\ell_2$ variant of \cref{eq:l1_ae}:
\begin{align}\label{eq:l12_ae}
 \min_{\mb W, \mb V} \; \frac{\norm{\wh{\mb X} - g_{\mb V} \circ f_{\mb W}  (\wh{\mb X})}_{1}}{\norm{\wh{\mb X} - g_{\mb V} \circ f_{\mb W}  (\wh{\mb X})}_{F}}.
  \quad (\text{$\ell_1/\ell_2$-RAE})
\end{align}
Empirically, this $\ell_1/\ell_2$-RAE consistently outperforms the $\ell_1$-RAE in several aspects, as we show in our experiments.


%% file: sections/Sec4_Exp.tex
\section{Experiments}\label{sec:experiments}
\begin{table}[!htpb]
\caption{The generalization performance of $\ell_1$-RAE (ours) and $\ell_1/\ell_2$-RAE (ours). The best PSNRs are colored as \textcolor{red}{red}; the best SSIMs are colored as \textcolor{blue}{blue}.}
\vspace{-7mm}
\label{tab:unseen}
\begin{center}
\setlength{\tabcolsep}{1mm}{
\begin{tabular}{c c c c c| c c c}
&
& 
\multicolumn{3}{c}{\scriptsize{PSNR $\uparrow$}} &
\multicolumn{3}{c}{\scriptsize{SSIM $\uparrow$}}
\\
\hline

\hline
\scriptsize{Dataset}

& \scriptsize{Corr.}

& \scriptsize{Baseline}
& \scriptsize{$\ell_1$-RAE}
& \scriptsize{$\ell_1/\ell_2$-RAE}

& \scriptsize{Baseline}
& \scriptsize{$\ell_1$-RAE}
& \scriptsize{$\ell_1/\ell_2$-RAE}
\\
\hline

\hline
\multirow{4}{*}{\rotatebox{90}{\scriptsize{MNIST}}}

&\scriptsize{10\%}

&\scriptsize{13.22}
&\scriptsize{24.40}
&\scriptsize{\textcolor{red}{24.32}}

&\scriptsize{0.59}
&\scriptsize{\textcolor{blue}{0.96}}
&\scriptsize{\textcolor{blue}{0.96}}

\\
\cline{2-8}

&\scriptsize{20\%}

&\scriptsize{10.19}
&\scriptsize{22.92}
&\scriptsize{\textcolor{red}{23.01}}

&\scriptsize{0.46}
&\scriptsize{\textcolor{blue}{0.94}}
&\scriptsize{\textcolor{blue}{0.94}}
\\
\cline{2-8}

&\scriptsize{30\%}

&\scriptsize{8.42}
&\scriptsize{\textcolor{red}{20.51}}
&\scriptsize{20.27}

&\scriptsize{0.38}
&\scriptsize{\textcolor{blue}{0.91}}
&\scriptsize{\textcolor{blue}{0.91}}
\\
\cline{2-8}

&\scriptsize{40\%}

&\scriptsize{7.17}
&\scriptsize{\textcolor{red}{16.47}}
&\scriptsize{15.28}

&\scriptsize{0.30}
&\scriptsize{\textcolor{blue}{0.80}}
&\scriptsize{0.75}
\\
\hline

\hline
\multirow{4}{*}{\rotatebox{90}{\scriptsize{CIFAR10}}}

&\scriptsize{10\%}

&\scriptsize{15.07}
&\scriptsize{36.33}
&\scriptsize{\textcolor{red}{38.15}}

&\scriptsize{0.47}
&\scriptsize{0.98}
&\scriptsize{\textcolor{blue}{0.99}}
\\
\cline{2-8}

&\scriptsize{20\%}

&\scriptsize{12.05}
&\scriptsize{31.20}
&\scriptsize{\textcolor{red}{36.27}}

&\scriptsize{0.31}
&\scriptsize{0.95}
&\scriptsize{\textcolor{blue}{0.98}}
\\
\cline{2-8}

&\scriptsize{30\%}

&\scriptsize{10.29}
&\scriptsize{27.36}
&\scriptsize{\textcolor{red}{31.84}}

&\scriptsize{0.22}
&\scriptsize{0.90}
&\scriptsize{\textcolor{blue}{0.96}}
\\
\cline{2-8}

&\scriptsize{40\%}

&\scriptsize{9.04}
&\scriptsize{22.69}
&\scriptsize{\textcolor{red}{24.43}}

&\scriptsize{0.16}
&\scriptsize{0.77}
&\scriptsize{\textcolor{blue}{0.88}}
\\
\hline

\hline
\multirow{4}{*}{\rotatebox{90}{\scriptsize{CIFAR100}}}

&\scriptsize{10\%}

&\scriptsize{14.95}
&\scriptsize{34.66}
&\scriptsize{\textcolor{red}{36.26}}

&\scriptsize{0.46}
&\scriptsize{\textcolor{blue}{0.98}}
&\scriptsize{\textcolor{blue}{0.98}}
\\
\cline{2-8}

&\scriptsize{20\%}

&\scriptsize{11.93}
&\scriptsize{30.74}
&\scriptsize{\textcolor{red}{35.78}}

&\scriptsize{0.29}
&\scriptsize{0.95}
&\scriptsize{\textcolor{blue}{0.98}}
\\
\cline{2-8}

&\scriptsize{30\%}

&\scriptsize{10.17}
&\scriptsize{26.84}
&\scriptsize{\textcolor{red}{31.19}}

&\scriptsize{0.21}
&\scriptsize{0.89}
&\scriptsize{\textcolor{blue}{0.95}}
\\
\cline{2-8}

&\scriptsize{40\%}

&\scriptsize{8.92}
&\scriptsize{22.80}
&\scriptsize{\textcolor{red}{25.28}}

&\scriptsize{0.15}
&\scriptsize{0.76}
&\scriptsize{\textcolor{blue}{0.87}}
\\
\hline

\hline
\end{tabular}
}
\end{center}
\vspace{-6mm}
\end{table}
\begin{table*}[t]
\caption{Results on the impact of the level of sample size on the performance. The ``g." inside the parenthesis means ``given''; the ``u." inside the parenthesis represents ``unseen".}
\label{tab:sample_size}
\vspace{-4mm}
\begin{center}
\setlength{\tabcolsep}{0.9mm}{
\begin{tabular}{l c c c c c c c| c c c c c c}
&
& 
\multicolumn{6}{c}{\scriptsize{$\ell_1$-RAE}} &
\multicolumn{6}{c}{\scriptsize{$\ell_1/\ell_2$-RAE}}
\\
\hline

\hline
\scriptsize{Corr.}

& \scriptsize{Exp.}

& \scriptsize{PSNR (g.)$\uparrow$}
& \scriptsize{PSNR (u.)$\uparrow$}
& \scriptsize{$\Delta$-PSNR$\downarrow$}
& \scriptsize{SSIM (g.)$\uparrow$}
& \scriptsize{SSIM (u.)$\uparrow$}
& \scriptsize{$\Delta$-SSIM$\downarrow$}

& \scriptsize{PSNR (g.)$\uparrow$}
& \scriptsize{PSNR (u.)$\uparrow$}
& \scriptsize{$\Delta$-PSNR$\downarrow$}
& \scriptsize{SSIM (g.)$\uparrow$}
& \scriptsize{SSIM (u.)$\uparrow$}
& \scriptsize{$\Delta$-SSIM$\downarrow$}
\\
\hline

\hline
\multirow{3}{*}{\rotatebox{90}{\scriptsize{10\%}}}

&\scriptsize{$\text{L}_{exp}$}

&\scriptsize{36.41}
&\scriptsize{36.38}
&\scriptsize{0.03}
&\scriptsize{0.98}
&\scriptsize{0.98}
&\scriptsize{0.00}

&\scriptsize{38.36}
&\scriptsize{38.31}
&\scriptsize{0.05}
&\scriptsize{0.99}
&\scriptsize{0.99}
&\scriptsize{0.00}
\\
\cline{2-14}

&\scriptsize{$\text{M}_{exp}$}

&\scriptsize{35.77}
&\scriptsize{35.74}
&\scriptsize{0.03}
&\scriptsize{0.98}
&\scriptsize{0.98}
&\scriptsize{0.00}

&\scriptsize{36.75}
&\scriptsize{36.68}
&\scriptsize{0.07}
&\scriptsize{0.99}
&\scriptsize{0.99}
&\scriptsize{0.00}
\\
\cline{2-14}

&\scriptsize{$\text{S}_{exp}$}
&\scriptsize{26.18}
&\scriptsize{21.88}
&\scriptsize{4.30}
&\scriptsize{0.87}
&\scriptsize{0.71}
&\scriptsize{0.16}

&\scriptsize{23.26}
&\scriptsize{19.58}
&\scriptsize{3.68}
&\scriptsize{0.81}
&\scriptsize{0.63}
&\scriptsize{0.18}
\\
\hline












\hline
\multirow{3}{*}{\rotatebox{90}{\scriptsize{30\%}}}

&\scriptsize{$\text{L}_{exp}$}

&\scriptsize{27.23}
&\scriptsize{27.21}
&\scriptsize{0.02}
&\scriptsize{0.90}
&\scriptsize{0.90}
&\scriptsize{0.00}

&\scriptsize{31.60}
&\scriptsize{31.57}
&\scriptsize{0.03}
&\scriptsize{0.96}
&\scriptsize{0.96}
&\scriptsize{0.00}
\\
\cline{2-14}

&\scriptsize{$\text{M}_{exp}$}

&\scriptsize{27.02}
&\scriptsize{26.99}
&\scriptsize{0.03}
&\scriptsize{0.90}
&\scriptsize{0.90}
&\scriptsize{0.00}

&\scriptsize{31.48}
&\scriptsize{31.44}
&\scriptsize{0.04}
&\scriptsize{0.96}
&\scriptsize{0.96}
&\scriptsize{0.00}
\\
\cline{2-14}

&\scriptsize{$\text{S}_{exp}$}
&\scriptsize{24.00}
&\scriptsize{18.93}
&\scriptsize{5.07}
&\scriptsize{0.81}
&\scriptsize{0.54}
&\scriptsize{0.27}

&\scriptsize{22.21}
&\scriptsize{17.45}
&\scriptsize{4.76}
&\scriptsize{0.76}
&\scriptsize{0.45}
&\scriptsize{0.31}
\\
\hline











\hline

\hline
\end{tabular}
}
\end{center}
\vspace{-8mm}
\end{table*}

\subsection{Datasets, Sparse Corruption, and Implementation}\label{subsec:datasets}
We choose three standard image datasets: MNIST, CIFAR-10, and CIFAR-100, reflecting increasing levels of variability and, hence, complexity of the underlying image manifolds. To simulate sparsely corrupted images, we add salt-and-pepper noise to clean images with \textit{corruption ratio} $p =\{10\%, 20\%, 30\%, 40\%\}$ to cover both benign and severe scenarios. To assess the quality of image recovery, we use PSNR~\cite{chan2005salt} and SSIM~\cite{wang2004image}. Our encoder consists of layers of strided convolution, and our decoder consists of layers of upsampling and convolution. We organize the layers into residual blocks~\cite{he2016deep} for both the encoder and decoder. For all models, we use the Adam optimizer with the learning rate $0.0001$.

\subsection{Collective Corruption Removal}\label{subsec:com_collective}

We take $60,000$, $50,000$, $50,000$ images from MNIST, CIFAR-10, and CIFAR-100, respectively, and then add sparse corruption to them with the aforementioned protocol. We stress that the corrupted images $\wh{\mb X}$ are the only data available and our goal is to restore the underlying clean images ${\mb X}$. We compare both $\ell_1$-RAE and $\ell_1/\ell_2$-RAE with RDA~\cite{zhou2017anomaly}, NRPCA~\cite{lyu2019manifold}, and RAE-CHA~\cite{chalapathy2017robust}. For these compared methods, we use their official implementations.

We calculate the average PSNR and SSIM of restorations by each method and also calculate these metrics on the corrupted images themselves to serve as our ``baseline". These numbers are presented in~\cref{tab:given}. We can make three observations: 1) on all datasets and across all corruption ratios, our $\ell_1$-RAE and $\ell_1/\ell_2$-RAE perform consistently better than the other methods, sometimes by a large margin;  2) $\ell_1/\ell_2$-RAE performs roughly comparably with $\ell_1$-RAE on MNIST, and it performs consistently better than $\ell_1$-RAE on CIFAR-10 and CIFAR-100, this might be due to the better approximation of $\ell_1/\ell_2$ to the $\ell_0$ norm;  3) the performance of each method drops as the corruption ratio grows, which is expected as the problem becomes harder.

\subsection{Generalizability to Unseen Noisy Images}\label{subsec:comp_dop}
In this experiment, \emph{we test if the learned manifold structures can be generalized to unseen noisy images effectively}. We take $10,000$ images from each dataset to construct unseen corrupted images. We note that there is no overlap between the images we used here and the images in~\cref{subsec:com_collective}. We only test the generalization performance of $\ell_1$-RAE (ours) and $\ell_1/\ell_2$-RAE (ours), as 1) $\ell_1$-RAE and $\ell_1/\ell_2$-RAE outperform RDA, NRPCA, and RAE-CHA by a considerable margin in the previous experiment, and 2) both RDA and NRPCA involve intensive iterative computation on new corrupted images. 
\cref{tab:unseen} summarizes the results. The remarkable thing to observe is that the performance on unseen data is on par with the performance as reported in Table~\ref{tab:given}, implying strong generalization and validity of the learned manifold structures.

\subsection{Impact of Sample Size}\label{subsec:sample_num}

Intuitively, we need sufficient samples to cover the manifolds reasonably densely in order to learn the manifold structures, especially when there is sparse corruption in the samples. In this section,~\emph{we quantitatively study how the varying sampling levels affect corruption-removal and generalization}.  We take CIFAR-10 and vary the level of the given samples from the training set: small ($S_{\mathrm{exp}}$) with $10, 000$ samples, medium ($M_{\mathrm{exp}}$) with $30, 000$ samples, and large ($L_{\mathrm{exp}}$) with $50, 000$ samples. Collective manifold learning and corruption removal is performed on these samples. Then, generalization of the learned model is tested on the $10,000$ unseen corrupted samples from \cref{subsec:comp_dop}. To measure the impact of the sample size on generalization, we also define the generalization gap, which measures the difference between the performance on given and unseen noisy images.

The results are summarized in~\cref{tab:sample_size}. The overall trend is consistent with our intuition: for a fixed corruption ratio, as the level of samples decreases, the corruption removal performance drops and the generalization gap is enlarged; with the same level of samples, both corruption removal and generalization performances drop as the corruption ratio increases.

%% file: sections/Sec6_Dis.tex
\section{Conclusion}\label{sec:discussion}
In this work, we have proposed arguably the simplest model for robust manifold learning against sparse corruption: $\ell_1$-RAE. To better handle sparsity, we have also proposed a variant $\ell_1/\ell_2$-RAE that empirically outperforms $\ell_1$-RAE. We have demonstrated that both perform effective manifold learning and corruption removal on given data and win competing methods by a considerable margin. The performance also generalizes almost perfectly to unseen samples with light computation. 

%% file: Main.bbl
\begin{thebibliography}{10}

\bibitem{HuberRonchetti2009Robust}
Peter~J. Huber and Elvezio~M. Ronchetti,
\newblock {\em Robust {Statistics}},
\newblock Wiley {Series} in {Probability} and {Statistics}. Jan. 2009.

\bibitem{candes2011robust}
Emmanuel~J. Cand{\`{e}}s, Xiaodong Li, Yi~Ma, and John Wright,
\newblock ``Robust principal component analysis?,''
\newblock {\em J. {ACM}}, vol. 58, no. 3, pp. 11:1--11:37, 2011.

\bibitem{chandrasekaran2011rank}
Venkat Chandrasekaran, Sujay Sanghavi, Pablo~A. Parrilo, and Alan~S. Willsky,
\newblock ``Rank-sparsity incoherence for matrix decomposition,''
\newblock {\em {SIAM} J. Optim.}, vol. 21, no. 2, pp. 572--596, 2011.

\bibitem{lu1998image}
Haw-Minn Lu, Yeshaiahu Fainman, and Robert Hecht-Nielsen,
\newblock ``Image manifolds,''
\newblock in {\em Applications of Artificial Neural Networks in Image
  Processing III}, 1998, vol. 3307, pp. 52--63.

\bibitem{belhumeur1998set}
Peter~N. Belhumeur and David~J. Kriegman,
\newblock ``What is the set of images of an object under all possible
  illumination conditions?,''
\newblock {\em Int. J. Comput. Vis.}, vol. 28, no. 3, pp. 245--260, 1998.

\bibitem{wang_early_2021}
Hengkang Wang, Taihui Li, Zhong Zhuang, Tiancong Chen, Hengyue Liang, and
  Ju~Sun,
\newblock ``Early stopping for deep image prior,''
\newblock {\em CoRR}, vol. abs/2112.06074, 2021.

\bibitem{li_self-validation:_2021}
Taihui Li, Zhong Zhuang, Hengyue Liang, Le~Peng, Hengkang Wang, and Ju~Sun,
\newblock ``Self-validation: Early stopping for single-instance deep generative
  priors,''
\newblock in {\em 32nd British Machine Vision Conference 2021, {BMVC} 2021},
  2021, p. 108.

\bibitem{li_impulse_2021}
Chun Li, Jian Li, and Ze~Luo,
\newblock ``An impulse noise removal model algorithm based on logarithmic image
  prior for medical image,''
\newblock {\em Signal Image Video Process.}, vol. 15, no. 6, pp. 1145--1152,
  2021.

\bibitem{yuan_ell_2019}
Ganzhao Yuan and Bernard Ghanem,
\newblock ``$\ell_0$tv: {A} sparse optimization method for impulse noise image
  restoration,''
\newblock {\em {IEEE} Trans. Pattern Anal. Mach. Intell.}, vol. 41, no. 2, pp.
  352--364, 2019.

\bibitem{adam_combined_2021}
Tarmizi Adam, Raveendran Paramesran, Mingming Yin, and Kuru Ratnavelu,
\newblock ``Combined higher order non-convex total variation with overlapping
  group sparsity for impulse noise removal,''
\newblock {\em Multim. Tools Appl.}, vol. 80, no. 12, pp. 18503--18530, 2021.

\bibitem{lyu2019manifold}
He~Lyu, Ningyu Sha, Shuyang Qin, Ming Yan, Yuying Xie, and Rongrong Wang,
\newblock ``Manifold denoising by nonlinear robust principal component
  analysis,''
\newblock in {\em Conference on Neural Information Processing Systems 2019,
  NeurIPS 2019}, 2019, pp. 13390--13400.

\bibitem{chalapathy2017robust}
Raghavendra Chalapathy, Aditya~Krishna Menon, and Sanjay Chawla,
\newblock ``Robust, deep and inductive anomaly detection,''
\newblock in {\em Machine Learning and Knowledge Discovery in Databases -
  European Conference, {ECML} {PKDD} 2017}, 2017, vol. 10534, pp. 36--51.

\bibitem{srebro2004maximum}
Nathan Srebro, Jason Rennie, and Tommi Jaakkola,
\newblock ``Maximum-margin matrix factorization,''
\newblock {\em Advances in neural information processing systems}, vol. 17,
  2004.

\bibitem{li2019non}
Qiuwei Li, Zhihui Zhu, and Gongguo Tang,
\newblock ``The non-convex geometry of low-rank matrix optimization,''
\newblock {\em Information and Inference: A Journal of the IMA}, vol. 8, no. 1,
  pp. 51--96, 2019.

\bibitem{zhou2017anomaly}
Chong Zhou and Randy~C. Paffenroth,
\newblock ``Anomaly detection with robust deep autoencoders,''
\newblock in {\em 23rd {ACM} {SIGKDD} International Conference on Knowledge
  Discovery and Data Mining}, 2017, pp. 665--674.

\bibitem{boyd2011distributed}
Stephen~P. Boyd, Neal Parikh, Eric Chu, Borja Peleato, and Jonathan Eckstein,
\newblock ``Distributed optimization and statistical learning via the
  alternating direction method of multipliers,''
\newblock {\em Found. Trends Mach. Learn.}, vol. 3, no. 1, pp. 1--122, 2011.

\bibitem{baldi1989neural}
Pierre Baldi and Kurt Hornik,
\newblock ``Neural networks and principal component analysis: Learning from
  examples without local minima,''
\newblock {\em Neural Networks}, vol. 2, no. 1, pp. 53--58, 1989.

\bibitem{HoyerNon}
Patrik~O. Hoyer,
\newblock ``Non-negative sparse coding,''
\newblock in {\em 12th {IEEE} Workshop on Neural Networks for Signal
  Processing, {NNSP} 2002}, 2002, pp. 557--565.

\bibitem{rahimi2019scale}
Yaghoub Rahimi, Chao Wang, Hongbo Dong, and Yifei Lou,
\newblock ``A scale-invariant approach for sparse signal recovery,''
\newblock {\em {SIAM} J. Sci. Comput.}, vol. 41, no. 6, pp. A3649--A3672, 2019.

\bibitem{zhuang_blind_2022}
Zhong Zhuang, Taihui Li, Hengkang Wang, and Ju~Sun,
\newblock ``Blind image deblurring with unknown kernel size and substantial
  noise,''
\newblock {\em CoRR}, vol. abs/2208.09483, 2022.

\bibitem{chan2005salt}
Raymond~H. Chan, Chung{-}Wa Ho, and Mila Nikolova,
\newblock ``Salt-and-pepper noise removal by median-type noise detectors and
  detail-preserving regularization,''
\newblock {\em {IEEE} Trans. Image Process.}, vol. 14, no. 10, pp. 1479--1485,
  2005.

\bibitem{wang2004image}
Zhou Wang, Alan~C. Bovik, Hamid~R. Sheikh, and Eero~P. Simoncelli,
\newblock ``Image quality assessment: from error visibility to structural
  similarity,''
\newblock {\em {IEEE} Trans. Image Process.}, vol. 13, no. 4, pp. 600--612,
  2004.

\bibitem{he2016deep}
Kaiming He, Xiangyu Zhang, Shaoqing Ren, and Jian Sun,
\newblock ``Deep residual learning for image recognition,''
\newblock in {\em 2016 {IEEE} Conference on Computer Vision and Pattern
  Recognition, {CVPR} 2016}, 2016, pp. 770--778.

\end{thebibliography}
